\documentclass[reprint,amsfonts,amsmath,amssymb,aps,pra,superscriptaddress,10pt]{revtex4-2}
\usepackage{graphicx}
\usepackage{sansmathfonts}
\usepackage[colorlinks=true,linkcolor=black,anchorcolor=black,citecolor=black,filecolor=black,menucolor=black,runcolor=black,urlcolor=black]{hyperref}

\begin{document}

% frontmatter
% {
\title{Quantum photonic neural networks in time}

\author{Ivanna M. Boras Vazquez}
\email{19imbv@queensu.ca}
\affiliation{Centre for Nanophotonics, Department of Physics, Engineering Physics \& Astronomy, 64 Bader Lane, Queen's University, Kingston, Ontario, Canada K7L 3N6}

\author{Jacob Ewaniuk}
\email{jacob.ewaniuk@queensu.ca}
\affiliation{Centre for Nanophotonics, Department of Physics, Engineering Physics \& Astronomy, 64 Bader Lane, Queen's University, Kingston, Ontario, Canada K7L 3N6}

\author{Nir Rotenberg}
\email{nir.rotenberg@queensu.ca}
\affiliation{Centre for Nanophotonics, Department of Physics, Engineering Physics \& Astronomy, 64 Bader Lane, Queen's University, Kingston, Ontario, Canada K7L 3N6}

\date{\today}
% }

\begin{abstract}
We introduce the architecture and timing algorithm to realize a time-bin-encoded quantum photonic neural network (QPNN): a reconfigurable nonlinear photonic circuit inspired by the brain and trained to process quantum information. Unlike the typical spatially-encoded QPNN, time-encoded networks require the same number of photonic elements (e.g. phase shifters or switches) regardless of their size or depth. Here, we present a model of such a network and show how to include imperfections such as losses, routing errors and most notably distinguishable photons. As an example, we train the QPNN to realize a controlled-NOT gate, based on a hypothetical ideal Kerr nonlinearity. We then extend our model to a realistic two-photon nonlinearity due to scattering from a single, semiconductor quantum dot coupled to a photonic waveguide. We show that, using this realistic nonlinearity, the QPNN can be trained to act as a Bell-state analyzer which operates with a fidelity of 0.96 and at a rate only limited by losses. We further show that time gating can raise this fidelity to over 0.99, while still maintaining an efficiency exceeding 0.9. Overall, this work lays a framework for the first QPNN encoded in time, and provides a clear path to the scaling of these networks.
\end{abstract}

\maketitle

\section{Introduction}
\begin{figure*}[ht]
\includegraphics[width= 14 cm]{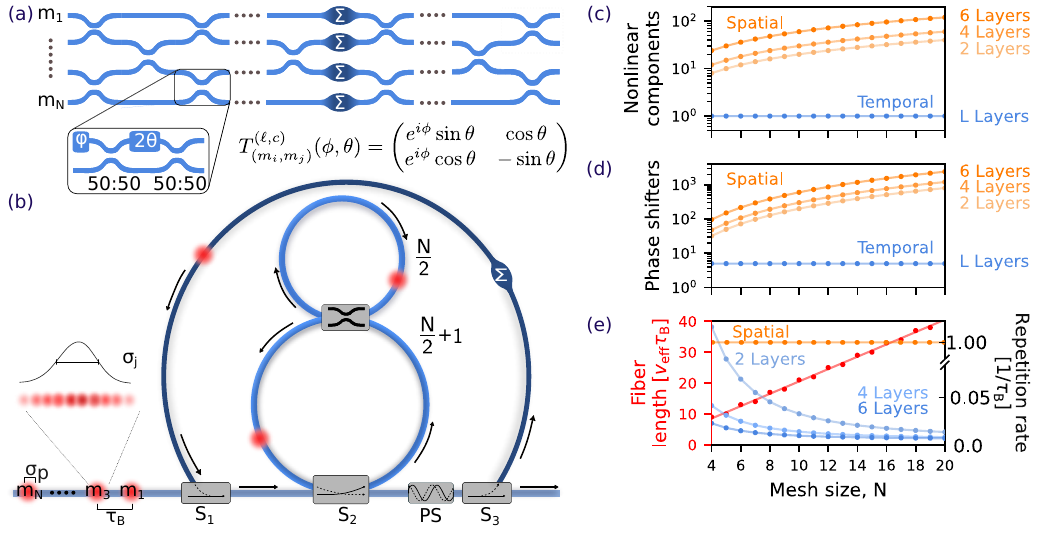}
\centering
\caption{Time- and space-encoded QPNNs. (a) The architecture of a spatially encoded $N \times  N$ QPNN comprised of meshes of linear interferometers (light blue) separated by single-site nonlinearities (dark blue, $\Sigma$). Each MZI contains two phase shifters ($\phi$, $2\theta$) and two $50:50$ directional couplers, as shown in the inset, performing a unitary transformation $T^{(\ell,c)}_{(m_{i},m_{j})}$ between modes $m_i$ and $m_j$. (b) The equivalent time-encoded QPNN architecture: two inner loops connected by a reconfigurable MZI realize the linear mesh, while the nonlinearity is part of the external loop. Different modes enter the QPNN successively through the bottom bus fiber, separated by a time-bin $\tau_\mathrm{B}$. Each photon pulse has a width $\sigma_\mathrm{p}$, and the time jitter of the pulses corresponds to a normal distribution with width $\sigma_\mathrm{j}$. Switches S\textsubscript{1}, S\textsubscript{2}, S\textsubscript{3} direct modes in and out of the linear and nonlinear loops, and PS is the output phase shifter. The required number of (c) nonlinear components and (d) phase shifters for the spatial (orange) and temporal (blue) QPNNs as a function of their size. (e) The requisite fiber length of the temporal QPNN (red, left axis), as well as the repetition rate at which the network operates normalized to $\tau_\mathrm{B}$, as a function of the network size, for the temporal (blue) and spatial (orange) networks (right axis).}
\label{fig:c1}
\end{figure*}

Discrete-variable quantum photonic neural networks (QPNNs) are brain-inspired light-based circuits that process individual photons. These networks are composed of linear interferometric meshes interspersed with nonlinear elements (denoted by $\Sigma$ in Figs.~\ref{fig:c1}a and b)
that enable them to learn to perform complex quantum operations such as generating Greenberger-Horne-Zeilinger states \cite{steinbrecher_quantum_2019}, acting as a Bell-state analyzer (BSA) \cite{ewaniuk_imperfect_2023}, logical processing on bosonic error-correcting codes \cite{Basani:25}, and even forming massive cluster states \cite{ewaniuk_large-scale_2025}. Since the nonlinear elements mediate photon-photon interactions, QPNNs perform these operations deterministically in contrast to linear quantum photonic circuits which are inherently probabilistic \cite{doi:10.1126/science.aab3642,Nielsen2025ProgrammableNonlinear}. In fact, even when constructed with imperfect photonic components, QPNNs can learn to overcome unbalanced photon loss, routing errors and even weak nonlinearities to achieve near-unity fidelities with operational rates limited only by loss \cite{ewaniuk_imperfect_2023}.

While undoubtedly powerful, QPNNs are also complex devices that have yet to be realized. Two major hurdles must be overcome. First, the number of active phase shifters and nonlinear elements grows rapidly with the network size (number of modes, $N$) and depth (number of layers, $L$) for the spatially-encoded QPNN (orange curves in Figs.~\ref{fig:c1}c and d), presenting an engineering challenge. More fundamentally, an appropriate nonlinear element that provides a photon-number-dependent phase shift without distorting the photon wavefunction has yet to be realized. Originally, QPNNs were modeled with an ideal optical Kerr nonlinearity \cite{steinbrecher_quantum_2019}, and while there have been significant advances in photonic resonator design \cite{PhysRevLett.124.160501,Ahn2022PhotonicInverseDesign, Mishra2025FanoNestedRing}, observed few-photon Kerr nonlinearity strengths in integrated devices remain five orders of magnitude below what is needed \cite{PhysRevA.73.062305,Venkataraman2013FewPhotonPhaseModulation}. Alternatively, quantum emitters such as semiconductor quantum dots (QDs) embedded in photonic structures are known to provide efficient few-photon nonlinearities of the required strength \cite{le_jeannic_dynamical_2022,Loo2012OpticalNonlinearity,Javadi2015SinglePhotonNonlinear}. Recent proposals envision using these emitters, either individually as part of a complex photonic system \cite{Basani:25, Kim2017HybridIntegration} or in arrays \cite{Tiurev2022HighFidelityCluster, Juska2013QDArray, Schrinski:22, Levy-Yeyati:25}, to create the requisite nonlinearity. While promising, these have yet to be demonstrated, much less at the scale required for large QPNNs (cf. Fig.~\ref{fig:c1}c).

In this work, we address both of these challenges by presenting a time-bin-encoded architecture for QPNNs, as shown in Fig.~\ref{fig:c1}b. By encoding each mode in time, rather than space, we show that it is possible to realize networks of any scale using a constant number of components, at the expense of an increased operation time (blue curves in Figs.~\ref{fig:c1}c-e); most notably, only one nonlinear element is required. We do this by developing and investigating network models that include realistic photonic imperfections, different nonlinearities and, crucially, partially distinguishable quantum photonic states. We first explain how our timing protocol can be applied to form any linear interferometric mesh, using a controlled-NOT (CNOT) quantum logic gate as an example. As a second step, we add the nonlinear component to form the full QPNN, and train the network to perform the same CNOT operation. In both cases, we study the effect of photon distinguishability, here due to time jitter, on the gate fidelity and operational rate, unveiling that unlike photonic component imperfections, distinguishability is a fundamental flaw which QPNNs cannot learn to correct. Finally, we present the results of a QPNN trained to realize a BSA, yet now using the nonlinearity induced by scattering from a two-level quantum emitter. Unlike the ideal Kerr nonlinearity, the emitter provides a strong nonlinear phase change at the expense of distorting the photon wavefunction in a way that reduces indistinguishability. Surprisingly, we find the network can learn around these distortions to act as a BSA with fidelities in excess of 0.96. We show that this can be further increased to 0.995 by gating the output in time, albeit with an operational rate reduced by 14\%. These results provide a path to viable large-scale QPNNs, comprised of few, realistic photonic components.

%%%%%%%%%%%%%%%%%%%%%%%%%%%%%%%%%%%%%%%%%%%%%%%%%%%%%

\section{Linear Networks in Time}
Two fiber loops can create a reconfigurable mesh in time \cite{clements_linear_2018, Yard2022TimeResource,PhysRevA.92.052319}, meaning that linear unitary transformations of arbitrary size can be formed from a single switch and a tunable Mach-Zehnder Interferometer (MZI), as shown in Fig.~\ref{fig:c1}b (light blue section). Briefly, individual photons are injected from the left in time intervals $\tau_{\mathrm{B}}$ and travel to switch $\mathrm{S}_2$ where they are routed into the larger, bottom fiber loop (which can support up to $N/2 + 1$ modes). The MZI can then be set to reroute the photon, either keeping it in the bottom loop or passing it to the top loop (which supports $N/2$ modes), or perform the two-mode unitary $T_{\left(m_i, m_j\right)}^{(\ell,c)}$ between modes $m_i$ and $m_j$ in column $c$ of the $\ell$\textsuperscript{th} mesh. Note that these unitaries are identical for both the spatially and temporally encoded networks, meaning that the setting of each phase shifter can be found in the usual way \cite{clements_optimal_2016}.

\begin{figure*}[htb]
\centering
\includegraphics[width=\textwidth]{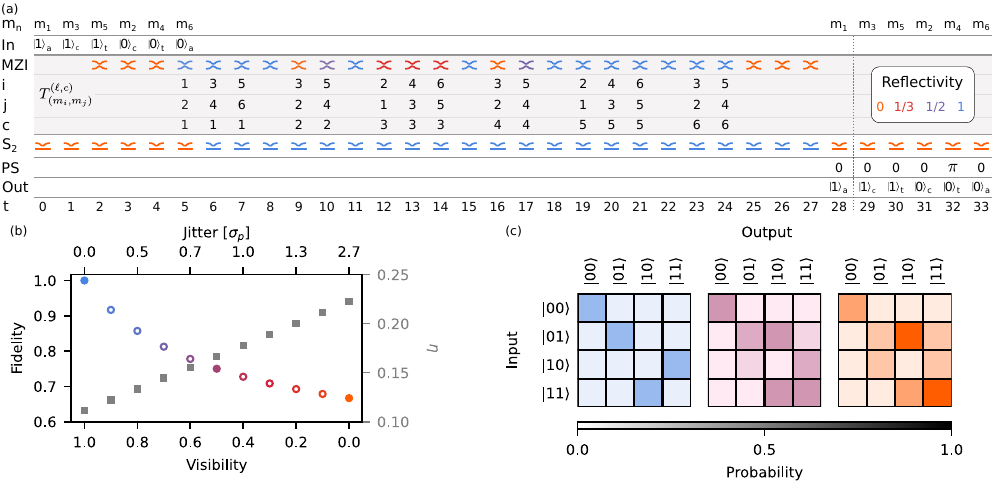}
\caption{A linear CNOT gate, in time. (a) CNOT gate timing algorithm, showing the reflectivity of the MZI, the state of switch S\textsubscript{2} and output phase shifter PS, at each timestep. For the timesteps where photons enter or leave the circuit, the modes ($m_n$) are labeled in correspondence with the equivalent spatial circuit (shown in Figs. S1 and S2 of the supplemental material). (b) The corresponding fidelity (colored dots, left axis) and efficiency (gray squares, right axis) as a function of visibility (or time jitter, top axis). (c) Hinton diagrams of the resulting CNOT operation for the visibilities corresponding to the solid markers in b ($V = 1, 0.5, 0$). Here, $|0\rangle$ and $|1\rangle$ denote the computational basis state for each input and output, and the output states are re-normalized to this basis.}
\label{fig:fig2}
\end{figure*}

Where our work diverges from previous implementations is in the timing protocol that, with an eye to multi-layer QPNNs, is designed to ensure the order in which photons (or, more precisely, the modes) are output matches the input. While precise details of the general timing protocol are available in Supplementary Information Sec.~S1, here we provide an example that illustrates the time-encoded linear network in operation. Specifically, we implement a 6-mode CNOT gate \cite{doi:10.1126/science.aab3642}, presenting the algorithm and results in Fig.~\ref{fig:fig2}.
We begin, in Fig.~\ref{fig:fig2}a, by showing the required settings for the MZI and $\mathrm{S}_2$ in each timestep $t_n = n\tau_{\mathrm{B}}$, while also denoting when the photons enter and leave the network.

As an example, consider the case where the control photon is in state $\left|1\right\rangle_\mathrm{c}$ (mode $m_3$) and the target photon is in state $\left|0\right\rangle_\mathrm{t}$ (mode $m_4$). The control photon enters the temporal circuit at time $t_1$, propagates through the bottom loop for two time-steps, and passes through the MZI at $t_3$. It then circles the top loop, returning to the MZI at time $t_6$. Separately, the target photon enters the bottom loop of the circuit at $t_4$, just in time to meet the control photon at the MZI. When they meet (at $t_6$), the MZI interferes the photons, performing $T_{(3,4)}^{(1,1)}$ (second MZI from the top in the first column of the mesh in Fig.~\ref{fig:c1}a). The rest of the MZI transformations are similarly achieved as the photons continue to circle the fiber loops, until time $t_{28}$, when the photons begin exiting the circuit. The full timing sequence, depicting the positions of each time-bin as they propagate through the circuit, can be found in Figs.~S1 and S2 of the Supplementary Information.

To quantify the performance of the quantum photonic network, we construct the transfer function of the circuit, $S$, working in a Fock basis that is extended to accommodate partial distinguishability between the photons. In fact, for imperfect photons, we can express the transfer function as $S=S_\mathrm{ind}\oplus S_\mathrm{dis}$, explicitly noting that the network processes indistinguishable and distinguishable photons differently (see Supplementary Information Sec.~2 for more details). Likewise, the density matrix representing any mixed input state can be written as $\rho_\mathrm{in}=V\rho_\mathrm{ind} + (1-V)\rho_\mathrm{dis}$, where the Hong-Ou-Mandel (HOM) visibility denotes the probability that the photons are indistinguishable \cite{PhysRevLett.59.2044,doi:10.1139/cjp-2023-0312}. The circuit operation results in an output state $\rho_\mathrm{out}=S\rho_\mathrm{in}S^\dagger$ that can be compared with the target $\rho_\mathrm{targ}$ to evaluate the fidelity $F$ (i.e. the chance the output is correct when a logical state is produced). We also calculate the operational rate of the network,
\begin{equation} \label{eq:rate}
    r = \frac{\eta(1-\alpha)}{n_t\tau_\mathrm{B}},
\end{equation}
where $\eta$ is the probability that the network performs its operation (i.e. results in a logical state, hereon referred to as efficiency), $n_t$ is the number of time-steps in the protocol, $\tau_\mathrm{B}$ is the time-bin duration, and $\alpha$ is the overall photon loss. In short, $\eta$ (and similarly, $r$) tells us how often the circuit works, and $F$ how well it works (see Supplementary Information Sec.~3 for full mathematical definitions).

Fig.~\ref{fig:fig2}b shows $F$ calculated for a controlled-NOT (CNOT) gate enacted on our time-encoded photonic network as a function of visibility. Since there is only one MZI in our circuit, it does not suffer from unbalanced losses or imperfect routing as do spatial networks \cite{doi:10.1126/science.aab3642}, so $F=1$ when $V=1$. Indeed, the reconstructed input-output table for this scenario, shown in blue in Fig.~\ref{fig:fig2}c, shows that an ideal CNOT has been implemented. We also plot $\eta$ in Fig.~\ref{fig:fig2}b, showing that in the ideal case this corresponds to the expected $1/9$ \cite{PhysRevA.65.062324}. Given that this protocol requires 28 timesteps (from when the first photon enters to when it exits), if we assume a 10 ns time bin duration (100 MHz repetition rate), then the operational rate for an ideal, lossless circuit would be nearly 400~kHz. Using state-of-the-art components \cite{aghaee_rad_scaling_2025}, $\alpha=0.36$, resulting in $r = 254$~kHz (see Supplementary Information Sec.~3 for loss modeling details).

If time jitter worsens and the photons become increasingly distinguishable, the fidelity decreases. As we observe in Fig.~\ref{fig:fig2}b, $F$ drops to 0.75 when $V=0.5$ and finally to 0.67 when $V = 0$. The corresponding Hinton diagrams, shown in purple and orange in Fig.~\ref{fig:fig2}c, respectively, increasingly diverge from the ideal (shown in blue). Interestingly, $\eta$ (and hence $r$) increases as $V$ drops, meaning that photons are more likely to be found in the logical basis (relative to the ideal case) but not in the correct state.

%%%%%%%%%%%%%%%%%%%%%%%%%%%%%%%%%%%%%%%%%%%%%%%%%%%%%

\section{QPNNs in Time}
The linear networks presented in the preceding section required only the two inner fiber loops of Fig.~\ref{fig:c1}b. Since our algorithm ensures that the modes enter and leave this circuit in the same order, extending this to form a QPNN simply requires the addition of one outer loop with an embedded nonlinearity $\Sigma$. Here, we start with a fictitious Kerr nonlinearity as is standard in the literature, which applies a nonlinear (i.e. photon-number-dependent) phase shift without distorting the pulses \cite{steinbrecher_quantum_2019}; in the following section, we relax this and show that the network also works using a realistic, quantum emitter-based nonlinearity.

\begin{figure}[ht]
\includegraphics[width=0.9\columnwidth]{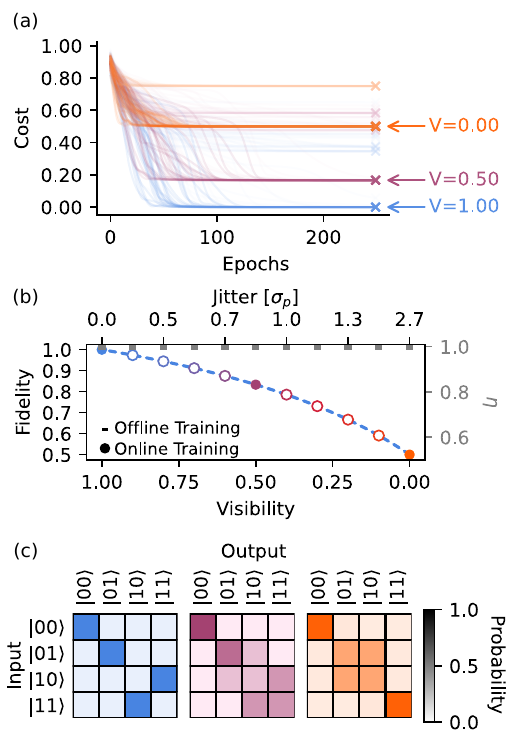}
\caption{Time-encoded 2-layer, 4-mode QPNN trained to perform a CNOT gate. (a) Training cost for 100 trials of 250 epochs each, for different visibilities. Successful training attempts clump together, resulting is overlapping curves that together appear darker. (b) Fidelity $F$ (colored, left axis) and efficiency $\eta$ (gray squares, right axis) as a function of visibility and time jitter. Circular markers denote online training while the dashed line shows offline training, which is done by taking the optimal solutions for $V=1$ and applying them to the system at different visibilities. (c) Hinton diagrams showing CNOT implementations at $V=1$, $0.5$ and $0$ (solid circles in b).}
\label{fig:fig3}
\end{figure}

We again construct a system transfer function, here including the nonlinearity (detailed in Supplementary Information Sec.~2), and use it to calculate the fidelity and operational rate. The addition of the nonlinearity enables us to train the QPNN \cite{ewaniuk_imperfect_2023}, which we do by minimizing the cost function (i.e. network error, the chance that an incorrect output is produced). We do so again for a CNOT gate, presenting the results from 100 training runs of two-layer networks, for different visibilities, in Fig.~\ref{fig:fig3}a.
Although not all training runs result in successfully learning to enact the CNOT operation (lighter regions, less curves), it is clear that many find the optimal solutions (darker regions, more curves).

The visibility-dependence of $F$ and $\eta$ for our time-encoded QPNNs is presented in Fig.~\ref{fig:fig3}b by the circle markers (online training), with the corresponding Hinton diagrams for $V=1$, $0.5$ and $0$ (corresponding to the data in panel a) shown in blue, purple and orange, respectively in Fig.~\ref{fig:fig3}c. We again find that $F=1$ for indistinguishable photons $\left(V = 1\right)$, and observe that this fidelity monotonically decreases to $F = 0.5$ when $V=0$. Interestingly, an identical behavior is observed for offline training (dashed blue line), where the optimal solution for $V=1$ is taken and then used with lower visibility inputs. This demonstrates that the network is unable to learn to compensate for distinguishability, in contrast to imperfections such as photon loss, routing errors and even sub-optimal nonlinearities \cite{ewaniuk_imperfect_2023}. In fact, the fidelity of the QPNN is more strongly affected than that of the linear network (c.f., Fig.~\ref{fig:fig2}b), which only decreases to 0.67.

The efficiency with which the QPNNs perform the CNOT remains at unity regardless of $V$ (right axis, grey square markers in Fig.~\ref{fig:fig3}b), in contrast to the linear network (Fig.~\ref{fig:fig3}b). That is, while in both cases $F$ decreases with $V$, the QPNN remains deterministic, always performing the desired operation. The QPNN is, however, larger than the linear circuit. For the example presented here, 58 timesteps were needed, meaning that with a 100 MHz repetition rate and in the lossless case, the CNOT gate can be performed at a 1.7~MHz rate (in contrast to the 0.4~MHz rate of the linear network). If state-of-the-art losses are again included, this drops to 741~kHz (compared to 254~kHz for the linear circuit). The trade-off is clear: the QPNN is deterministic in nature, always performing its target operation, but this comes at a cost of more time-steps and loss.

%%%%%%%%%%%%%%%%%%%%%%%%%%%%%%%%%%%%%%%%%%%%%%%%%%%%%

\section{QPNNs with Quantum Dot Nonlinearities}
All components of a time- (or space-) encoded QPNN exist, except for a Kerr-based nonlinear element. In fact, at the few-photon level, integrated Kerr nonlinear strengths are currently five orders of magnitude weaker than what is required for a QPNN \cite{PhysRevA.73.062305, Venkataraman2013FewPhotonPhaseModulation}. While there are several promising proposals for realizing the requisite nonlinearity, they typically require ultra-high quality \cite{doi:10.1126/science.1193968,Lin:17} or dynamical cavities \cite{PhysRevLett.124.160501, PhysRevA.101.042322,Heuck2020ControlledPhase}, complex nanophotonic circuits \cite{Basani:25, RevModPhys.87.347} or multiple quantum emitters acting in unison \cite{PhysRevA.105.063501,Tudela2015SubwavelengthQED}, all of which have yet to be demonstrated. Here, we show that passive scattering from a single, well-coupled quantum emitter (which we take to be a quantum dot, QD) can, in fact, produce a nonlinearity sufficient for high-fidelity quantum operations.

\begin{figure}[tbp]
\centering
\includegraphics[width=0.9\columnwidth]{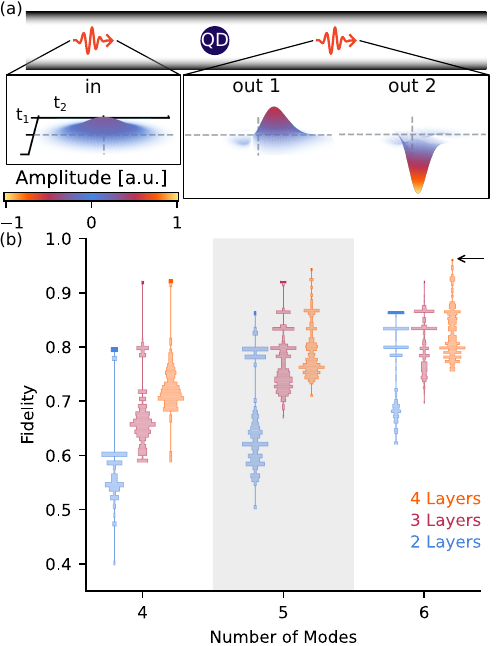}
\caption{Time-encoded QPNN trained to act as a BSA, using the nonlinearity of a waveguide-coupled QD. (a) Schematic of a QD in a waveguide, showing how an input one- (two-) photon wavefunction (in time) is altered after scattering. (b) Fidelity distributions for QPNNs of different sizes, each trained to act as BSAs. The width of each box corresponds to the number of QPNNs that achieved the fidelity at which it is found. The arrow denotes the optimal QPNN with $N=6$ and $L=4$, which reached a fidelity of 0.96, and is studied further in Fig.~\ref{fig:fig5}.}
\label{fig:fig4}
\end{figure}

Strong, two-photon nonlinearities have recently been observed with QDs coupled to integrated waveguides, albeit for symmetric QD-waveguide coupling where nearly all single-photons are reflected \cite{le_jeannic_dynamical_2022}. For chirally-coupled QDs \cite{lodahl_chiral_2017}, theory predicts near-lossless nonlinear scattering \cite{PhysRevA.91.043845} as we show in Fig.~\ref{fig:fig4}a;
here, for on-resonance scattering, where $\omega_\mathrm{p} = \omega_{\mathrm{QD}}$, and when the photon pulse width is the same as the QD lifetime, where $\sigma_{\mathrm{p}} = \tau_{\mathrm{QD}}$ (see Supplementary Information Sec.~4 for details on the scattering theory). In this case, the nonlinear phase imparted by the QD is exactly that of the ideal Kerr nonlinearity: a $\pi$ phase difference between the single- and two-photon responses. Unfortunately, the scattering also significantly distorts the photon wavefunction, delaying a photon that scatters alone and creating spectrotemporal correlations between photons that scatter together. As a consequence, the single- and two-photon components become partially distinguishable with respect to each other, reducing the fidelity of subsequent linear layers in a similar way as time jitter, previously.

There are important quantum operations that are typically paired with a subsequent measurement and are hence largely insensitive to distortions of the photon wavefunction. Notably, the BSA, a key component of one-way \cite{PhysRevX.10.021071} and two-way quantum repeaters \cite{Mantri2025QuantumRepeaters}, which projects quantum states onto the maximally entangled basis, is one such operation. When implemented with linear optics (and no ancillary photons), a BSA has a $\eta=0.5$ efficiency with $F=1$ in the ideal case \cite{PhysRevA.59.3295}. In contrast, QPNN-based BSAs with fictitious Kerr nonlinearities can reach both $\eta=1$ and $F=1$ in the ideal case \cite{steinbrecher_quantum_2019}, with the rate only decreasing due to losses \cite{ewaniuk_imperfect_2023}.

Here, we train QPNNs to act as BSAs based on the passive QD scattering nonlinearity. We train the networks as usual using the MZI phase shifts, yet now add the QD lifetime, $\tau_{\mathrm{QD}}$ (with one value for the entire network), and the detuning between the photons and the QD, $\Delta=\omega_\mathrm{QD}-\omega_{\mathrm{p}}$ (with one value per layer), as trainable parameters. In contrast to typical BSAs, we do not map each of the four maximally entangled Bell states $\left\{\left|\Phi^{\pm}\right\rangle , \, \left|\Psi^{\pm}\right\rangle  \right\}$ to a logical state (e.g. $\left|01\right\rangle$ or $\left|11\right\rangle$) directly, but rather assign a random mapping at the beginning of each training run, insisting only that the target output states should result in distinct measurement outcomes, all of which have one photon in some mode $m_i$ while the other photon must be in a different mode $m_j$. This is particularly important as we add ancillary modes (but not photons) to the circuit, increasing the dimensionality for the mapping, thus allowing the network to search for higher fidelity solutions. This is an attractive approach with time-encoded circuits, as these additional modes do not require additional resources (neither photonic components nor photons), only an increase in the time per operation. More details on training for this model can be found in Supplementary Information Sec.~5.

We show the BSA fidelity achieved from training QPNNs with different numbers of layers and modes in Fig.~\ref{fig:fig4}b. Here, we show the entire distribution of trained networks for each case, with the width of each line denoting the number of QPNNs that achieved that fidelity, similar to a histogram. In each case, we darken the lines at the top of the distribution to help identify the highest fidelity QPNNs, noting that in all darkened regions $\eta =1$. For example, with no ancillary modes (i.e. $N=4$), the smallest QPNNs ($L=2$) reach $F=0.79$. Since $\eta=1$, this means that the QPNN produces a logical output 100\% of the time, at double the rate of a linear circuit, yet falsely classifies the state $21\%$ of the time. Adding a third and fourth layer increases this maximum fidelity to $0.92$ and $0.93$, respectively, as the added resources enable the QPNN to find better solutions. Note that, in all cases, most training runs fail and even for the largest ($N=4$ and $L=4$) networks, only about 1 out of 100 are able to find the optimal solution.

To improve network fidelity, we introduce additional modes, which both increase the number of trainable parameters as well as the number of distinct output mode combinations that can be mapped to. For example, adding a single ancillary mode $\left(N=5\right)$ increases the number of distinct two-mode combinations from 6 to 10, while a sixth mode increases this further to 15. As we observe in Fig.~\ref{fig:fig4}b, increasing the network size in this manner increases the fidelity, which reaches 0.96 for $L=4$, $N=6$ QPNNs.

\begin{figure}[tbp]
\centering
\includegraphics[width=0.9\columnwidth]{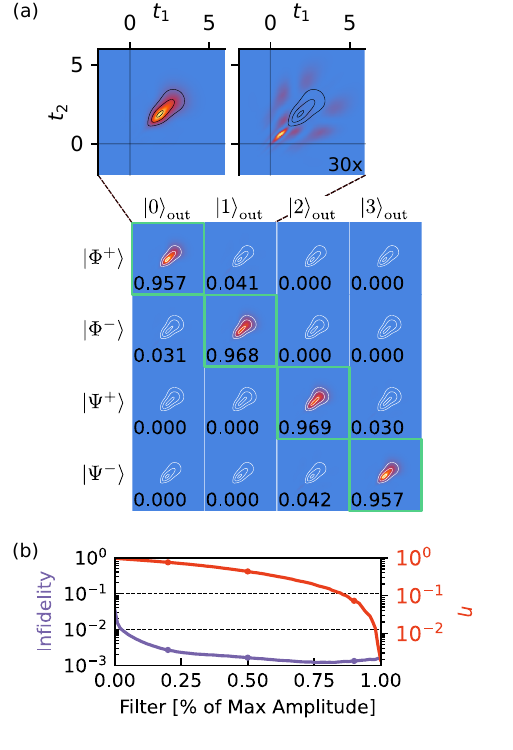}
\caption{Approaching unity fidelity for a QD-based QPNN trained as a BSA. (a) The two-photon temporal wavefunction in each logical output defined for the BSA while training the QPNN shown with an arrow in Fig.~\ref{fig:fig4}b. The fraction of the probability distribution found in each output is annotated, with those that signal a correct result outlined in green. In each bin, we mark exemplary time filters calculated as contours at 0.2, 0.5, and 0.9 of the maximum amplitude for the distribution in the correct output. For clarity, we expand these in the inset, showing states $\left|0\right\rangle_\mathrm{out}$ (correct state) and $\left|1\right\rangle_\mathrm{out}$ (false positive; amplified by a factor of 30) with filters. (b) Infidelity (purple) and efficiency (red) of the BSA as a function of the filter drawn at a percentage of the maximum value. The points on each curve correspond to the filters shown in a.}
\label{fig:fig5}
\end{figure}

We show the time-dependent two-photon probability distribution in each target state, for each input, for the highest fidelity QPNN (marked with an arrow in Fig.~\ref{fig:fig4}b) in Fig.~\ref{fig:fig5}a.
Since this circuit has ancillary modes, each distribution is comprised of multiple, distinct two-photon output states that are together assigned to the same Bell-state; for example, $\left|0\right\rangle_\mathrm{out}$ consists of the measurement outcomes $\left\{(m_1, m_2), (m_2, m_4), (m_3, m_6)\right\}$ (see Supplementary Information Sec.~5 for the full mapping). Ideally, each input would map with unity probability to a single target output (denoted by a green frame), yet this is clearly not the case. For example, $\left|\Phi^{+}\right\rangle $ only maps to $\left|0\right\rangle _{\mathrm{out}}$ with a probability of 0.957, while there is a probability of 0.041 of finding this state in $\left|1\right\rangle _{\mathrm{out}}$ and $<0.002$ in the remaining output states. We find a similar distribution for the mapping of the other Bell-states. 

Interestingly, a closer inspection of the output wavefunctions presented in Fig.~\ref{fig:fig5}a shows that there is little temporal overlap between the component in the desired output and the rest. To show this clearly, we zoom in on $\left|0\right\rangle _{\mathrm{out}}$ and $\left|1\right\rangle _{\mathrm{out}}$ for the $\left|\Phi^{+}\right\rangle $ input. Here, we show exemplary time filters (contours) chosen to encompass areas of high probability in $\left|0\right\rangle _{\mathrm{out}}$. The 3 exemplary filters shown are drawn at 0.2, 0.5 and 0.9 of the maximum amplitude in $\left|0\right\rangle _{\mathrm{out}}$. Note that, for these cases and assuming $\tau_{\mathrm{QD}}=1$~ns, we require 1.05, 0.66 and 0.21~ns time windows, well above the few ps resolution of superconducting nanowire single photon detectors \cite{kerman2013readout,Gourgues:19} (see Supplementary Information Sec.~6 for further details). That is, with a modest sacrifice to $\eta$, we are able to substantially increase the operational fidelity of the QD-based BSA. In fact, as summarized in Fig.~\ref{fig:fig5}b, we can reach $F=0.99$ with only a drop to $\eta=0.93$, and $F=0.995$ for $\eta=0.864$. For a time bin duration of 10 ns, the latter corresponds to an ideal operation rate of 745~kHz that, realistically, will only be limited by circuit losses.

%%%%%%%%%%%%%%%%%%%%%%%%%%%%%%%%%%%%%%%%%%%%%%%%%%%%%

\section{Discussion}
In summary, we have mapped an architecture capable of near-deterministic  complex quantum operations, here a BSA ($F=0.995$ at $\eta=0.864$), using a QPNN based on the passive scattering from a single, chirally-coupled quantum emitter. To do so, we present a time-encoding architecture for the QPNN, along with a timing algorithm that allows for the realization of arbitrarily large circuits at the cost of operation rate, but not resources (such as photons, photonic elements such as phase-shifters, or quantum emitters). Since our architecture relies on only a single MZI, losses are inherently balanced, meaning that they do not decrease $F$, only $\eta$. Encouragingly, the losses of quantum photonic components are rapidly decreasing and, due to industrial investment \cite{aghaee_rad_scaling_2025, Alexander2025Nature}, are likely to continue on this trend.

In this work, we focused on the passive scattering of two photons from a single, asymmetrically-coupled QD to implement the requisite nonlinearity. We did so because two-photon scattering from a QD has been observed \cite{le_jeannic_dynamical_2022} and quantified \cite{le_jeannic_experimental_2021} in integrated platforms, albeit in the symmetric configuration. However, other quantum emitters may instead be used, including atoms coupled to a nanofiber \cite{PhysRevLett.104.203603}, Rydberg atoms (where three-photon scattering has been observed) \cite{doi:10.1126/science.aao7293}, or defects in diamond \cite{PhysRevX.14.041013} or silicon \cite{Hollenbach:20}. Similarly, QPNNs based on quantum emitters need not be limited to 2-photon operations, as was the case for the BSA considered here. More complex operations, such as the generation of GHZ \cite{PhysRevLett.132.130604,Pont2024} or cluster states \cite{ewaniuk_large-scale_2025} may well be possible if suitable multi-photon interactions can be identified. Existing models of multi-photon scattering \cite{PhysRevA.91.043845, arranz_regidor_modeling_2021}, coupled with recent demonstrations of platforms supporting multi-photon scattering experiments \cite{PhysRevA.82.063816}, enable exploration of this regime of nonlinear quantum optics. These, together with the architecture and results presented here, provide a path to the next generation of neuromorphic quantum photonic devices.

%%%%%%%%%%%%%%%%%%%%%%%%%%%%%%%%%%%%%%%%%%%%%%%%%%%%%

\begin{acknowledgments}
This work was supported by the Canadian Foundation for Innovation (CFI), the National Science and Engineering Research Council (NSERC), and Queen’s University. The authors gratefully acknowledge insightful discussions with Dr. Bhavin Shastri.
\end{acknowledgments}

%%%%%%%%%%%%%%%%%%%%%%%%%%%%%%%%%%%%%%%%%%%%%%%%%%%%%

\section*{Data Availability}
The code used to produce the findings of this study is the quotonic package, available at https://github.com/jewaniuk/quotonic/. The data produced during this work will be openly available in the Borealis repository of the Queen’s University Dataverse.

%%%%%%%%%%%%%%%%%%%%%%%%%%%%%%%%%%%%%%%%%%%%%%%%%%%%%

\section*{Author Contributions}
I.M.B.V. derived the timing protocol for time-encoded QPNN circuits, then prepared the full model, including time jitter, alongside J.E. J.E. investigated the use of a quantum dot as a QPNN nonlinearity. I.M.B.V. performed all simulations and analysis, with supervision from N.R. The results were discussed by all authors, who also shared the writing and editing responsibilities for the manuscript.

% references
% {
\bibliography{ref}
% }

\end{document}

% frontmatter
% {
\title{Supplementary Information for ``Quantum photonic neural networks in time''}

\author{Ivanna M. Boras Vazquez}
\email{19imbv@queensu.ca}
\affiliation{Centre for Nanophotonics, Department of Physics, Engineering Physics \& Astronomy, 64 Bader Lane, Queen's University, Kingston, Ontario, Canada K7L 3N6}

\author{Jacob Ewaniuk}
\email{jacob.ewaniuk@queensu.ca}
\affiliation{Centre for Nanophotonics, Department of Physics, Engineering Physics \& Astronomy, 64 Bader Lane, Queen's University, Kingston, Ontario, Canada K7L 3N6}

\author{Nir Rotenberg}
\email{nir.rotenberg@queensu.ca}
\affiliation{Centre for Nanophotonics, Department of Physics, Engineering Physics \& Astronomy, 64 Bader Lane, Queen's University, Kingston, Ontario, Canada K7L 3N6}

\date{\today}
% }

\maketitle

\section{Timing Protocol for Time-Encoded Circuits}
\begin{figure*}[p]
\includegraphics[width=14cm]{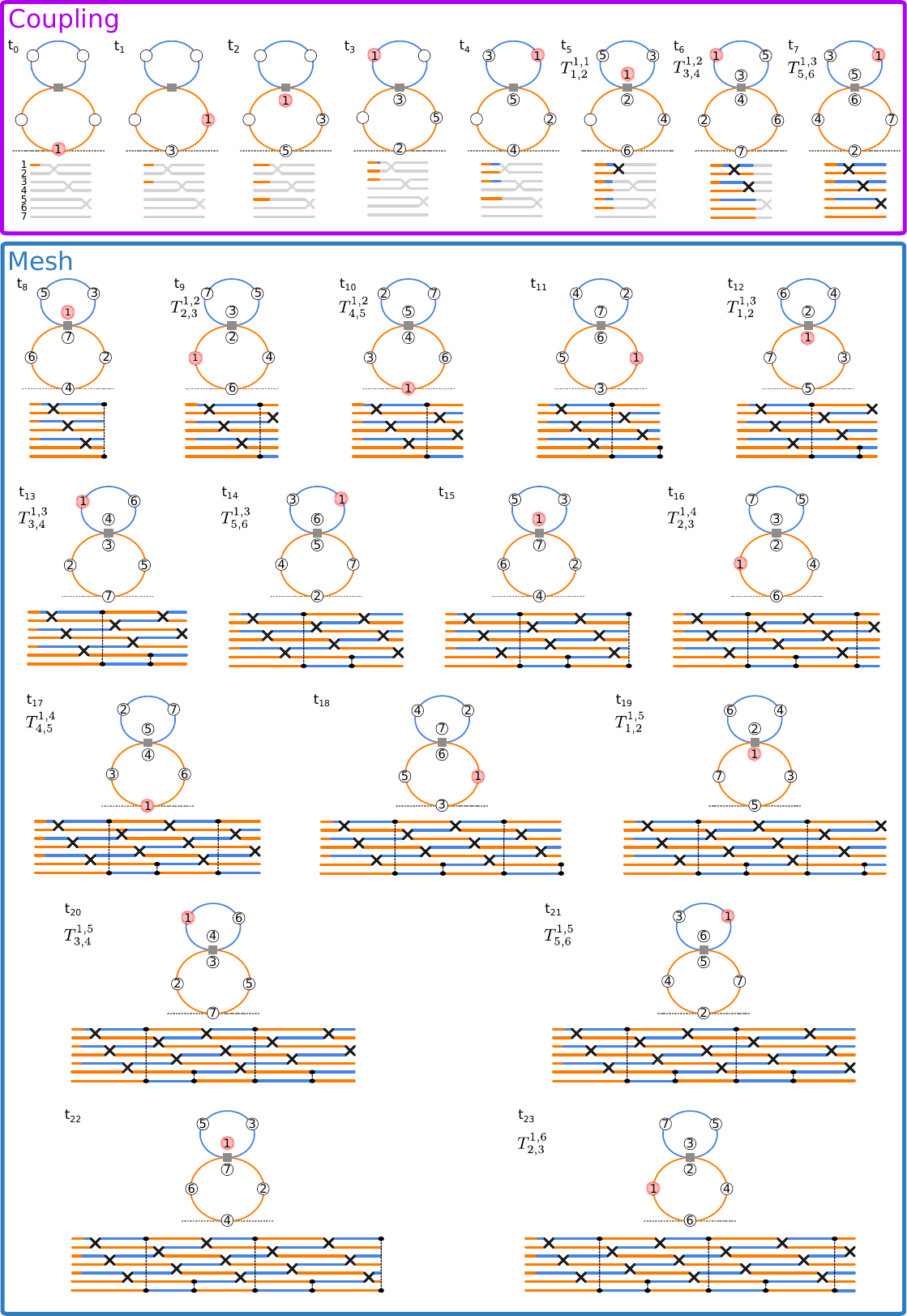}
\centering
\caption{Step by step time-bin example with its spatial counterpart below it for a CNOT gate. The in/out bus fiber is denoted by the grey dashed line. Each step (starting from $\mathrm{t_0}$) shows how the modes move through the circuit, with its corresponding spatial mesh being built underneath it. The grey colour coding of the spatial mesh in the coupling section dentoes what the mesh will look like once the coupling steps are complete. We colour the spatial modes in orange to show that the mode is found in the bottom loop, while the blue colour coding denotes the mode being in the top loop. Each black cross in the spatial mesh corresponds to an MZI, and is denoted by a grey square in the time-bin architecture. For relevant timesteps the corresponding transformation $T^{(\ell,c)}_{(m_i,m_j)}$ is shown (see Fig. 2a in main text for more details). The first mode is highlighted in red for readability.  }
\label{fig:mesh}
\end{figure*}

\begin{figure*}[p]
\includegraphics[width=14cm]{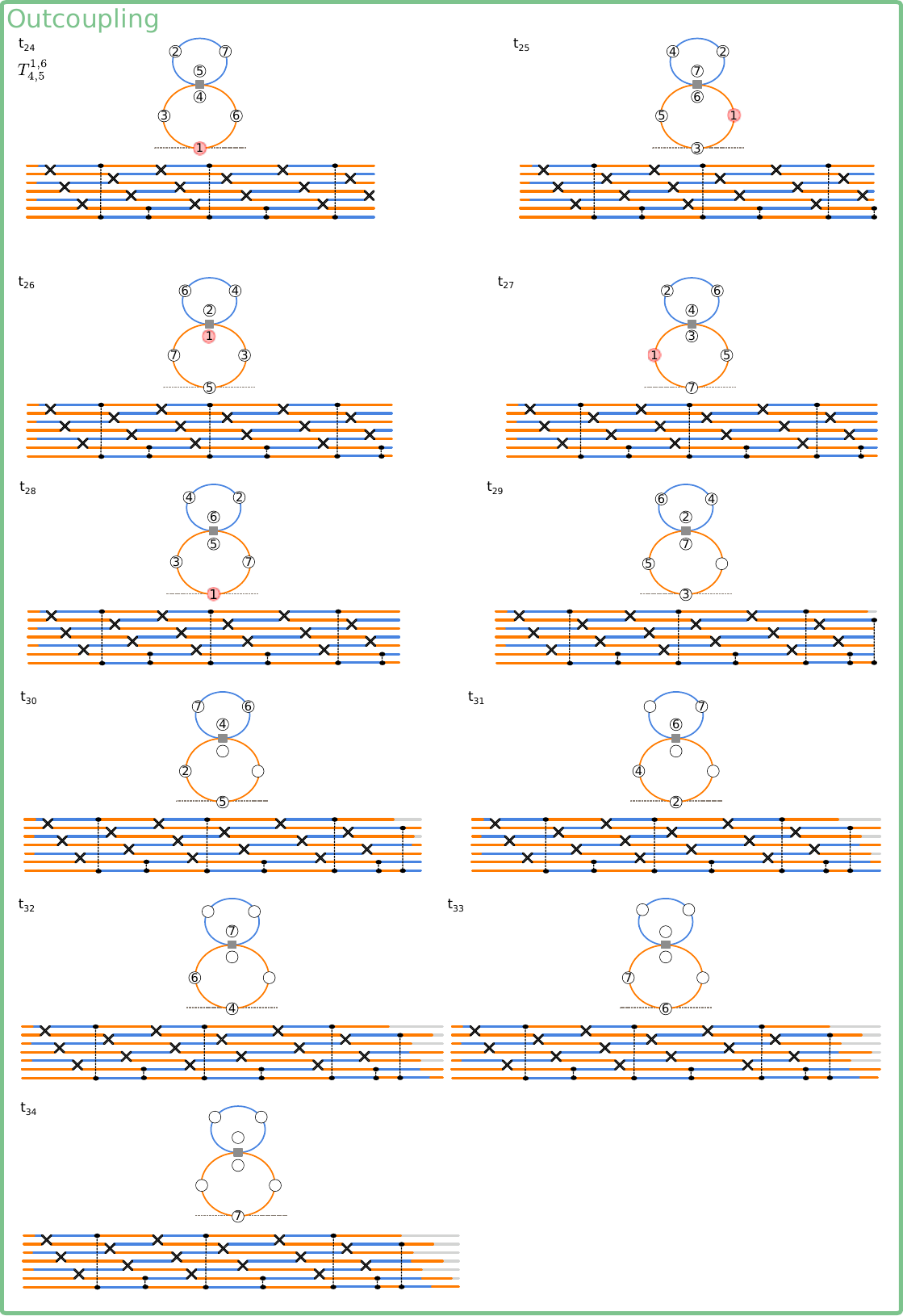}
\centering
\caption{Outcoupling of mesh, continuation from Fig.~\ref{fig:mesh}. Here there is a re-ordering of modes to ensure they leave the circuit in the same way they entered (odd modes first, followed by even modes). At ${t_{28}}$ the first mode is coupled out of the circuit, marking the end of its trajectory in the mesh. The grey colour coding in the spatial mesh denotes modes that have completed their trajectory and left the mesh.}
\label{fig:mesh2}
\end{figure*}

The time-bin circuit is composed of two loops of different sizes, the top loop scaling by $\lceil N/2 \rceil$ and the bottom loop scaling by $\lceil N/2+1 \rceil$. For example, if $N=5$ then the top loop would be 3 time-bin widths long while the bottom loop would be 4. At time ${t_{0}}$, the first mode enters, followed by all the rest of the odd modes at each time-bin. Once all odd modes are coupled in, the even modes follow. An extra (empty) ancillary mode $(N+1)$ is added to even-$N$ meshes after all modes have been coupled into the system. Figs. \ref{fig:mesh} and \ref{fig:mesh2} illustrate how the modes propagate through a circuit for a $N=6$ mesh in both the spatial and time-encoded architecture. Here, we follow the transformations done for a CNOT gate, with the transformations $T^{(\ell,c)}_{(m_i,m_j)}$ corresponding directly to Fig. 2a in the text. The coupling of modes, as well as the mesh building, is shown in Fig. \ref{fig:mesh}, while the outcoupling of modes is shown in Fig. \ref{fig:mesh2}. Each timestep shows where the modes are found in the time-bin architecture, as well as what the spatial mesh looks like at that time. For ease of following the mode's trajectories, mode 1 has been highlighted in red. Each black dot and dashed line corresponds to a loop swap, and no transformation $T^{(\ell,c)}_{(m_i,m_j)}$ being applied.
The general algorithm for an even mesh is generally as follows, for ease we will follow the example given in Figs. \ref{fig:mesh} and \ref{fig:mesh2}. The modes $(1-6)$ are numbered beside the first spatial mesh.
\begin{enumerate}
    \item Starting at $t_{0}$, couple all odd modes $m_{\mathrm{odd}}$ into circuit, all $m_{\mathrm{odd}}$ (as seen in Fig. 2a in the main text) couple to the top loop. Once all $m_{\mathrm{odd}}$ have been coupled, begin coupling even modes $m_{\mathrm{even}}$ into circuit. In the example being followed, this occurs at $t_{3}$.
    \item Begin interference when last $m_{\mathrm{even}}$ is added (leaving room for an empty ancillary mode $m_{{N+1}}$ at end), $m_{\mathrm{odd}}$ route to top loop.
    \item Once first column of interferences is complete, swap ancillary mode with $m_{\mathrm{1}}$. In the example, this occurs at $t_{8}$ and is denoted by the dashed black line on the spatial mesh. 
    \item Continue interference, swapping spots between $m_{\mathrm{odd}}$ and $m_{\mathrm{even}}$. Once the second column is done, $m_{{N}}$ and $m_{{N+1}}$ swap. In the example, this occurs at $t_{11}$.
    \item Continue interference, swapping spots between $m_{\mathrm{odd}}$ and $m_{\mathrm{even}}$.
    \item Repeat steps 3 and 4 $N/2$ times. 
    \item Begin out coupling, where in the example this begins at $t_{24}$, in Fig.~\ref{fig:mesh2}. Here, all $m_{\mathrm{odd}}$ are kept in the bottom loop, as illustrated in steps $t_{26}$ to $t_{28}$, where $m_{\mathrm{1}}$ is coupled out. The next timestep ($t_{29}$ in example), the $m_{\mathrm{even}}$ modes are coupled into the bottom loop so they can leave the mesh. All $m_{\mathrm{odd}}$ are coupled out, followed suit by $m_{\mathrm{even}}$.
\end{enumerate}
The general algorithm for an odd mesh is similar, with the only distinction being that it lacks the empty ancillary $m_{{N+1}}$ mode. The general algorithm is as follows:
\begin{enumerate}
    \item Input all odd modes ($m_{\mathrm{odd}}$) into circuit, route all odd modes to top loop. 
    \item Input all even modes ($m_{\mathrm{even}}$) into circuit.
    \item Begin interference when last $m_{\mathrm{even}}$ is added, $m_{\mathrm{odd}}$ route to top loop.
    \item Complete all mode interferences.
    \item Swap mode $N$ with $m_{\mathrm{1}}$.
    \item Continue interference, swapping spots between $m_{\mathrm{odd}}$ and $m_{\mathrm{even}}$.
    \item Mode $m_{\mathrm{N-1}}$ and $m_{\mathrm{N}}$ swap.
    \item Continue interference, swapping spots between $m_{\mathrm{odd}}$ and $m_{\mathrm{even}}$.
    \item Repeat steps 5 and 6 $N$ times, always followed by steps 7 and 8.
    \item Begin outcoupling sequence.
\end{enumerate}
The necessary number of steps for a mode of the linear layer to complete its entire progression through the circuit is given by 
\begin{align}
    \chi=\frac{1}{2}(N+1)(N+2)+b,
    \label{eq:fullcricuit}
\end{align}
where $N$ denotes the number of modes in the circuit and $b$ is a buffer applied for the nonlinearity. In the example, this corresponds to step $t_{28}$, when mode $m_{\mathrm{1}}$ reaches the output switch. In the main text in Fig. 2a this is the last step before the grey dashed line.  Since at this point the switch is set to complete transmission, as is the MZI, the next linear mesh can start immediately after out coupling $m_{\mathrm{1}}$ out (in this case, $b=0$ since Fig. 2a deals with a linear circuit). In our case, when considering the QPNN, a buffer of 1 time-bin length is enough to complete the circuit. 

%%%%%%%%%%%%%%%%%%%%%%%%%%%%%%%%%%%%%%%%%%%%%%%%%%%%%

\section{QPNN Operation on Partially Distinguishable Photons due to Time Jitter}

The model used for treating partially distinguishable photons closely follows the one derived in Ref.~\cite{PhysRevA.98.043839}. In order to compute our input state, we first employ a change of basis using the Schur--Weyl transform. The transform enables us to express these states in terms of both symmetric and asymmetric combinations. Doing this (as explained in Ref.~\cite{PhysRevA.98.043839}) allows us to write our state in two parts, so we can define how much of our state is indistinguishable.
For two photons entering a two-mode interferometer, each photon has a \emph{System} (mode) and a \emph{Label} (distinguishability) degree of freedom. Schur--Weyl duality allows us to express two-photon states according to their permutation symmetry, corresponding to the irreducible representations (irreps) of $U(2)$. The symmetric (triplet) and antisymmetric (singlet) basis states are:
\begin{align}
|xx\rangle, \quad 
\frac{|xy\rangle + |yx\rangle}{\sqrt{2}}, \quad 
|yy\rangle, \quad 
\text{and} \quad 
\frac{|xy\rangle - |yx\rangle}{\sqrt{2}}.
\end{align}
In this case, indistinguishable photons occupy only the symmetric subspace, while distinguishable photons have components in both. Tracing out the label degree of freedom yields a reduced System density matrix that is block-diagonal, each block corresponding to an irrep. The unitary transformation in this basis is block-diagonal and given by
\begin{equation}
U^{(2)} = U \otimes U = 
\text{per}(U) \oplus \text{det}(U),
\end{equation}
where the symmetric block (permanent) describes bosonic interference and the antisymmetric block (determinant) corresponds to fermionic behavior. Thus, the Schur--Weyl basis naturally separates indistinguishable and distinguishable photon dynamics. Using this property we are able to construct the system function $S$. Each layer of the QPNN consists of a linear optical transformation $U$ (performed by the MZI mesh) followed by a nonlinear operation $\Sigma$, except for the final layer, where it ends with a linear transformation. Since the circuit acts on the discrete degrees of freedom (label) we are able to write the system function as 
\begin{equation}
    S =
    U({\phi}_L, {\theta}_L, {\delta}_L)
    \prod_{\ell=1}^{L-1}
    \Sigma \,
    U({\phi}_\ell, {\theta}_\ell,{\delta}_\ell),
\end{equation}
where $L$ denotes the total number of layers in the QPNN, $\mathrm{\phi}$ and $\mathrm{\theta}$ are the phase shifts applied by the MZI and $\mathrm{\delta}$ is the output phase shift (PS in Fig. 1). The index $\ell \in \{1,\dots,L\}$ labels the individual layers, with $\ell = 1$ corresponding to the first layer and $\ell = L$ the final layer. 
Before applying the system function, we first construct the input state. Partial distinguishability is described in the continuous degrees of freedom (label), and its effect can be quantified by comparing the reduced input density matrix $\rho_{\mathrm{in}}$ to the ideal indistinguishable input state.
We begin by constructing the full input state $\rho_{\mathrm{in}}$ and tracing out the continuous degrees of freedom, which leaves us a density matrix that acts only on the discrete modes,
\begin{align}
    \rho_{\mathrm{in}} = \int d\omega_{1} \int d\omega_{2}
    \left\langle \omega_{1}, \omega_{2} \middle| \mathrm{in} \right\rangle
    \left\langle \mathrm{in} \middle| \omega_{1}, \omega_{2} \right\rangle.
\end{align}
We define the ideal input density matrix as
\begin{align}
\rho_{\mathrm{in}}^{(\mathrm{ideal})} = |\psi\rangle \langle \psi|,
\end{align}
where $|\psi\rangle$ is the corresponding indistinguishable two-photon state.
The effect of partial distinguishability on the input state can be quantified by comparing the reduced input density matrix $\rho_{\mathrm{in}}$ to the ideal indistinguishable input state. The difference between the ideal and actual input states is computed using the input fidelity, given by
\begin{align}
F_{\mathrm{in}} = \left( \mathrm{Tr} \sqrt{ \sqrt{\rho_{\mathrm{in}}^{(\mathrm{ideal})}} \, \rho_{\mathrm{in}} \, \sqrt{\rho_{\mathrm{in}}^{(\mathrm{ideal})}} } \right)^2.
\end{align}
Since the ideal state is pure, this simplifies to
\begin{align}
F_{\mathrm{in}} = \langle \psi | \rho_{\mathrm{in}} | \psi \rangle,
\end{align}
 directly measuring the overlap between the actual input state and the ideal (indistinguishable) state.
This fidelity is therefore directly related to the fraction of the input state that is indistinguishable. Using the HOM visibility $V$ re-normalized between $0$ and $1$ (following Ref.~\cite{PhysRevLett.103.053601}), this can be simply expressed as
\begin{align}
F_{\mathrm{in}} = \frac{1 + V}{2}.
\end{align}
This shows that the visibility determines the overlap of the input state with the indistinguishable components. We then propagate the state through the QPNN, and have an output density matrix which contains a mixture of indistinguishable and distinguishable contributions,
\begin{align}
\rho_{\mathrm{out}} =
V \, S \, \rho_{\mathrm{in}}^{(\mathrm{ind})} \, S^\dagger
+
(1 - V) \, S_{\mathrm{linear}} \, \rho_{\mathrm{in}}^{(\mathrm{dist})} \, S_{\mathrm{linear}}^\dagger,
\end{align}
This shows that the full effect of the circuit on partially distinguishable photons can be computed entirely from the reduced density matrix $\rho_{\mathrm{in}}$, without tracking the continuous (label) degrees of freedom, allowing us to disregard the distinguishable part of our state. When dealing with the QPNN, this is important, as the nonlinearity requires that both photons arrive at the same time. The parts that are offset in their overlap (distinguishable bits) are unable to trigger a nonlinear response, thus the regular system function does not apply to them. This is a crucial fact, as it allows us to simplify our output state as defined in Eq. S10, with the full system function acting on the indistinguishable state $\rho_{\mathrm{in}}^{(\mathrm{ind})} $ and only the linear system function (without the nonlinearity) acting on our distinguishable state $\rho_{\mathrm{in}}^{(\mathrm{dist})}$.

\begin{figure}[ht]
\includegraphics[width=0.5\columnwidth]{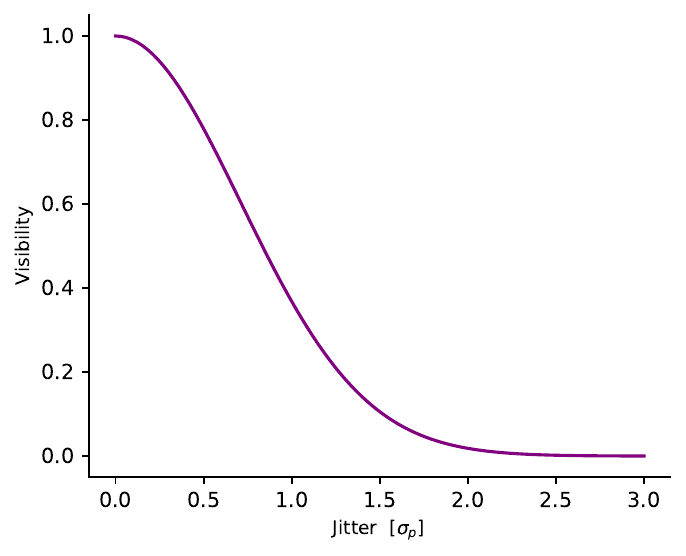}
\centering
\caption{HOM visibility as a function of time jitter, $\sigma_\mathrm{j}$, in units of the photon width $\mathrm{\sigma_{p}}$, obtained by sampling a Gaussian distribution centered at zero with width $\sigma_\mathrm{j}$ 200 times, and averaging the samples. For a given temporal offset sample, the HOM visibility is calculated following Ref.~\cite{doi:10.1139/cjp-2023-0312}.}
\label{fig:visvtj}
\end{figure}

Finally, time jitter is modeled using a Gaussian distribution of temporal offsets between the pair of photons. The mean of this distribution is zero, while the full width at half maximum represents the jitter, $\sigma_\mathrm{j}$. For a given amount of time jitter, the corresponding average HOM visibility can be evaluated by sampling from this Gaussian distribution in many iterations, calculating the HOM visibility for a specific temporal offset in each, and taking the average of all the samples (see Fig.~\ref{fig:visvtj} as a reference).

%%%%%%%%%%%%%%%%%%%%%%%%%%%%%%%%%%%%%%%%%%%%%%%%%%%%%

\section{Cost, Loss and Efficiency of the QPNN}
The cost function, evaluated for a $k^{\mathrm{th}}$ input-target pair, is found by comparing the difference between the output state $\rho_{\mathrm{out}}$ and the target state $\rho_{\mathrm{targ}}$ and is defined by
\begin{equation}
    C^{(k)}
    = 1-\frac{1}{F^{(k)}_{\mathrm{in}}}
    \left(
        \mathrm{Tr}
        \sqrt{
            \sqrt{\rho^{(k)}_{\mathrm{targ}}}
            \, \rho^{(k)}_{\mathrm{out}} \,
            \sqrt{\rho^{(k)}_{\mathrm{targ}}}
        }
    \right)^2.
\end{equation}
Here, we scale inversely by $F_{\mathrm{in}}$ to quantify the QPNN’s performance when the input photons are partially distinguishable. In other words, this allows us to exclude the time jitter source from the evaluation and find the intrinsic performance of the QPNN independent of the input state distinguishability.
For a training set of size $K$, the cost is averaged over all input-target pairs
\begin{align} \label{eq:Cavg}
    C_{\mathrm{avg}} &=  \, \frac{\mathrm{(1-\alpha)}}{K} \sum_{k=1}^K C^{(k)},
\end{align}
where  $\mathrm{(1-\alpha)}$ is the transmissivity. The total transmissivity is calculated by taking into account all sources of loss, and converting them from dB to linear (fractional) loss. The losses considered are due to the MZI ($\alpha_{
\mathrm{mzi}}$), phase shifter ($\alpha_{\mathrm{ps}}$), switch ($\alpha_{\mathrm{switch}}$), chip coupling ($\alpha_{\mathrm{chip}}$). 
 The number of times each mode traverses a component per linear layer is related to the number of modes ($N$) by
\begin{align}
\Omega_{\mathrm{mzi}} &= N + 2 - (N \bmod 2), \\
\Omega_{\mathrm{switch}} &= \left\lfloor \frac{1}{2}(N - 4) \right\rfloor + 6,
\end{align}
where $\Omega_{\mathrm{mzi}}$ are the number of times the MZI is traversed (per mode) and similarly for the switch we have $\Omega_{\mathrm{switch}}$.
The fiber length per linear layer is
\begin{align}
L_{\mathrm{fiber}}^{(\mathrm{lin})} = \frac{1}{2}(N+1)(N+2)\,\tau_{B}\,v,
\end{align}
where $v$ is the velocity of the light through the fiber dependent on the material's index of refraction (here we take the group index of 1.46, from SMF-28 optical fiber) and $\tau_{B}$ is the time bin length. We can use this to calculate the fractional loss for the fiber and define it as $\alpha_{\mathrm{fiber}}$. The total transmissivity per linear layer, $T_{\mathrm{lin}}$, is then calculated as
\begin{align}
T_{\mathrm{lin}} &= 
(1 - \alpha_{\mathrm{mzi}})^{\Omega_{\mathrm{mzi}}}
(1 - \alpha_{\mathrm{switch}})^{\Omega_{\mathrm{switch}}}
(1 - \alpha_{\mathrm{ps}})
(1 - \alpha_{\mathrm{fiber}}^{(\mathrm{lin})})
(1 - \alpha_{\mathrm{chip}})^{2\left(\Omega_{\mathrm{switch}} - 1\right)}.
\end{align}
For the nonlinear part of the QPNN, the fiber length is given by
\begin{align}
L_{\mathrm{fiber}}^{(\mathrm{nl})} = b \cdot \tau_{B} \cdot v,
\end{align}
where the buffer, $b$ is how many time-bins we expect the modes to be in the nonlinear part of the circuit, with fiber loss $\alpha_{\mathrm{fiber}}^{(\mathrm{nl})} $. The total transmissivity over L layers follows the same multiplication procedure, this time multiplying all the linear layers, and their nonlinear part. The value of $\mathrm{\alpha}$ in our case was calculated using state of the art components from \cite{aghaee_rad_scaling_2025}. The values used for loss yielded  $\alpha=0.36$ (where $\alpha$ indicates the total loss) for the linear CNOT circuit, and  $\alpha=0.87$ for the nonlinear 4 layer, 6 mode BSA.

We now define the efficiency ($\eta$) as the probability that the network performs its operation. The efficiency is given by
\begin{equation}
    \eta = \sum_{| i \rangle \epsilon \mathrm{CB}} \langle i | \rho_{\mathrm{out}} | i \rangle,
\end{equation}
where we sum over a set of computational basis states CB, and each $|i\rangle$ corresponds to a valid output configuration. In other words, $\eta$ is obtained by summing the diagonal elements (populations) of $\rho_{\mathrm{out}}$ restricted to the computational subspace. This quantity therefore represents the total probability that the photons remain in the desired modes and are successfully detected.
The fidelity, $F$, of our system is then defined as
\begin{equation}
    F = \frac{1}{\eta}
    \sum_{k=1}^K\frac{1}{F_\mathrm{in}^{(k)}}\left(
        \mathrm{Tr}
        \sqrt{
            \sqrt{\rho^{(k)}_{\mathrm{targ}}}
            \, \rho^{(k)}_{\mathrm{out}} \,
            \sqrt{\rho^{(k)}_{\mathrm{targ}}}
        }
    \right)^2, 
\end{equation}
which by comparison with Eq.~\ref{eq:Cavg} is the same as $F = (1 - C_\mathrm{avg}) / \eta(1-\alpha)$. The fidelity is conditioned on the production of a logical state at the circuit output, meaning as a measure it is independent of the efficiency $\eta$ and the loss $
\alpha$.
%%%%%%%%%%%%%%%%%%%%%%%%%%%%%%%%%%%%%%%%%%%%%%%%%%%%%

\section{Chiral Scattering from a Two-Level Quantum Emitter} \label{sec:scattering}
To model the chiral scattering of single- and two-photon Fock states from a two-level quantum emitter, we follow the input-output scattering matrix formalism explored extensively in the literature \cite{Fan:10, Shen:07, Mahmoodian:20}, which we briefly outline here for reference. Consider a two-level system (2LS) that is chirally-coupled to a quasi-infinite one-dimensional waveguide that has an approximately linear dispersion relation over the near-resonant frequencies of interest for this analysis. The Hamiltonian for this system is given by
\begin{equation} \label{eq:hamiltonian}
    \hat{H} = \frac{1}{2}\hbar\Delta\hat{\sigma}_z + \int_{-\infty}^\infty d\omega \hbar\omega\hat{a}^\dagger(\omega)\hat{a}(\omega) + \int_{-\infty}^\infty d\omega\frac{\hbar V}{\sqrt{2\pi v_g}}\left(\hat{a}^\dagger(\omega)\hat{\sigma}_- + \hat{\sigma}_+\hat{a}(\omega)\right),
\end{equation}
where $\Delta \equiv \omega_\mathrm{QD} - \omega_\mathrm{p}$ is the detuning between the transition frequency $\omega_\mathrm{QD}$ of the 2LS and the center frequency $\omega_\mathrm{p}$ of the waveguide dispersion (i.e. the photonic part), $\hat{\sigma}_z \equiv \hat{\sigma}_+\hat{\sigma}_- - \hat{\sigma}_-\hat{\sigma}_+$ where $\hat{\sigma}_\pm$ are the raising and lowering operators for the 2LS, $\hat{a}^\dag(\omega)$, $\hat{a}(\omega)$ are the creation and annihilation operators for photons of frequency $\omega$, $v_g$ is the group velocity, and $V$ is the coupling strength between the 2LS and the waveguide modes \cite{Fan:10}. The three terms of the Hamiltonian correspond to the energy levels of the 2LS, the photons in the waveguide, and the interaction between them which is treated under the dipole, rotating wave, and Markov approximations as are typically valid in experimental conditions \cite{le_jeannic_experimental_2021, le_jeannic_dynamical_2022}. For a QPNN, it is not explicitly necessary to describe the dynamics of the 2LS if we can extract the response that it imprints on the photons long after the interaction. Therefore, input and output operators are defined which create photonic Fock states in the limits $t \to -\infty$ and $t \to \infty$, respectively (i.e. long before and long after the interaction). Defining the scattering matrix $\Sigma$ as the operator which relates the scattering states (those created by the input and output operators) to the free photonic states (those created by the photonic operators of the Hamiltonian in Eq.~\ref{eq:hamiltonian}), the matrix elements of $\Sigma$ then describe how the response of the 2LS affects the output photons much after the interaction. In Ref.~\cite{Fan:10}, the matrix elements of the single-photon scattering matrix are derived as
\begin{equation} \label{eq:scattering1}
    \left\langle\omega'\right|\Sigma\left|\omega\right\rangle = t(\omega)\delta(\omega - \omega'),
\end{equation}
where $\left|\omega\right\rangle = \hat{a}^\dagger(\omega)\left|0\right\rangle$ with $\left|0\right\rangle$ representing the vacuum state, and
\begin{equation} \label{eq:tomega}
    t(\omega) \equiv \frac{\omega - \Delta - \frac{i}{2\tau_\mathrm{QD}}}{\omega - \Delta + \frac{i}{2\tau_\mathrm{QD}}}
\end{equation}
is the complex single-photon transmission coefficient where $\tau_\mathrm{QD}$ is the emission lifetime of the 2LS. Also in Ref.~\cite{Fan:10}, the matrix elements of the two-photon scattering matrix are derived as
\begin{align} \label{eq:scattering2}
    \left\langle\omega_1',\omega_2'\right|\Sigma\left|\omega_1,\omega_2\right\rangle &= \frac{1}{2}t(\omega_1)t(\omega_2)\left(\delta(\omega_1 - \omega_1')\delta(\omega_2 - \omega_2') + \delta(\omega_1 - \omega_2')\delta(\omega_2 - \omega_1')\right)\nonumber \\
    &\quad+ i\frac{\sqrt{\Gamma}}{2\pi}s(\omega_1')s(\omega_2')(s(\omega_1) + s(\omega_2))\delta(\omega_1 + \omega_2 - \omega_1' - \omega_2'),
\end{align}
where $\left|\omega_1,\omega_2\right\rangle = \frac{1}{\sqrt{2}}\hat{a}^\dag(\omega_1)\hat{a}^\dag(\omega_2)\left|0\right\rangle$, and
\begin{equation} \label{eq:somega}
    s(\omega) \equiv \frac{1}{\sqrt{\tau_\mathrm{QD}}}\frac{1}{\omega - \Delta + \frac{i}{2\tau_\mathrm{QD}}},
\end{equation}
which the authors of Ref.~\cite{Fan:10} claim is a measure of the 2LS excitation by a single photon of frequency $\omega$. While these matrix elements are explicitly defined for monochromatic photonic Fock states, we can use the fact that photons of finite wavepacket width are superpositions of these states weighted by a spectral amplitude, or equivalently, a wavefunction defined in the continuous frequency basis. At this point, we note that working in the frequency basis is merely a mathematical choice given the form of the scattering matrix elements. The photons can equivalently be described in time by simply transforming to the continuous time basis via a Fourier transform. In this work, we perform the numerical analysis in the frequency basis, where the scattering matrix is easier to model, then perform Fourier transforms as necessary afterward when investigating the results. In the frequency basis, the input single-photon states are defined as
\begin{equation} \label{eq:in1}
    \left|\mathrm{in}\right\rangle_1 = \int_{-\infty}^\infty d\omega \psi_\mathrm{in}(\omega)\left|\omega\right\rangle.
\end{equation}
Applying the scattering matrix, then inserting the identity operator followed by Eq.~\ref{eq:scattering1}, the output state from the single-photon interaction is given by
\begin{equation} \label{eq:out1}
    \left|\mathrm{out}\right\rangle_1 = \int_{-\infty}^\infty d\omega t(\omega)\psi_\mathrm{in}(\omega)\left|\omega\right\rangle.
\end{equation}
Following an analogous procedure, given the two-photon analog of Eq.~\ref{eq:in1} as the input state, the output state from the two-photon interaction is derived as
\begin{align} \label{eq:out2}
    \left|\mathrm{out}\right\rangle_2 = &\int_{-\infty}^\infty d\omega_1\int_{-\infty}^\infty d\omega_2 \Bigg[t(\omega_1)t(\omega_2)\psi_\mathrm{in}(\omega_1)\psi_\mathrm{in}(\omega_2) + \frac{i}{2\pi\sqrt{\tau_\mathrm{QD}}}s(\omega_1)s(\omega_2)\nonumber \\ &\times \int_{-\infty}^\infty dp \psi_\mathrm{in}(\omega_1 + \omega_2 - p)\psi_\mathrm{in}(p)\left(s(\omega_1 + \omega_2 - p) + s(p)\right)\Bigg]\left|\omega_1, \omega_2\right\rangle.
\end{align}
In this work, we consider input photons with a Gaussian envelope in time, implying that they must have a corresponding Gaussian spectral amplitude,
\begin{equation} \label{eq:gaussian}
    \psi_\mathrm{in}(\omega) = \left(\frac{\sigma_\mathrm{p}^2}{2\pi}\right)^{\frac{1}{4}}\exp{\left(-\frac{\sigma_\mathrm{p}^2}{4}(\omega - \omega_\mathrm{p})^2\right)},
\end{equation}
where $\sigma_\mathrm{p}$ is the photon pulse width in the time domain. This wavefunction is necessarily normalized, following the condition $\int_{-\infty}^\infty d\omega\left|\psi_\mathrm{in}(\omega)\right|^2 = 1$.

In Fig.~4a of the main text, we consider the case where $\sigma_\mathrm{p} = \tau_\mathrm{QD}$ and $\omega_\mathrm{p}=\omega_\mathrm{QD}$ ($\Delta = 0$) as an example. Here, we plot the amplitudes of the two-photon wavefunctions at the input (i.e. product of Eq.~\ref{eq:in1}, starting with initially uncorrelated Gaussian photons), and the output, both for the case where the photons scatter separately (i.e. product of Eq.~\ref{eq:out1}, as if the photons sequentially interact with the QD from separate time-bin modes) and that where the photons scatter together (Eq.~\ref{eq:out2}). For this particular condition, all of these two-photon wavefunctions are purely real such that we can plot them without regard for a complex phase. This allows us to visually identify the nonlinear $\pi$ phase difference between the single-photon response and two-photon response (i.e. in the two-photon response the wavefunction is negative).

%%%%%%%%%%%%%%%%%%%%%%%%%%%%%%%%%%%%%%%%%%%%%%%%%%%%%

\section{Training a QPNN with QD Nonlinearities to act as a BSA}
% outline how the model builds from the qd scattering theory to do the full qpnn simulation (follow phd promotion and candidacy reports)
% explain how the training function works and how outputs get assigned to each bell state
% show the full input-output mapping for the example discussed in Fig 5
% show the full wavefunction diagrams
It is clear from the form of Eqs.~\ref{eq:out1} and \ref{eq:out2} that we must explicitly consider photon wavepacket distortions while modeling QPNNs with nonlinearities based on coherent chiral scattering from two-level quantum emitters. While typical QPNN models, including that which we consider in Sec.~3 on the main text, use Fock states with photons that are only defined by their discrete spatial or temporal mode (i.e. $\left|s_1, s_2\right\rangle = \frac{1}{\sqrt{2}}\hat{a}^\dag_{s_1}\hat{a}^\dag_{s_2}\left|0\right\rangle$ where each $s_i$ is used as a label for the degree of freedom that encodes information), the extended model merely extends the Fock basis to a continuous one by also considering the frequency (or equivalently time) degree of freedom (i.e. $\left|\omega_1s_1, \omega_2s_2\right\rangle = \frac{1}{\sqrt{2}}\hat{a}^\dag_{s_1}(\omega_1)\hat{a}^\dag_{s_2}(\omega_2)\left|0\right\rangle$). The response of each linear layer is broadband relative to the spectral width of the photons, thus, they still only transform the photons' discrete degrees of freedom, acting essentially the same as in the previous model. The only difference is that now, rather than transforming an input which has a complex probability amplitude attached to each of the discrete optical mode basis states, the input instead has a complex wavefunction attached to the basis states.

Though the nonlinearities are modeled in a significantly different manner from the previous QPNN model (as detailed in Sec.~\ref{sec:scattering}), the new procedure follows exactly from the previous one. Each nonlinear element (i.e. quantum emitter) modulates the input wavefunctions according to the input-output scattering matrix formalism by simply choosing the appropriate scattering matrix depending on whether the corresponding optical mode basis state has two photons in the same mode or not. Computationally, this extension involves discretizing the infinite continuous basis. By adding this to the model, it is now inefficient to compute a general transfer function $S$ for the QPNN. Instead, the transformation enacted by the QPNN is computed and applied depending on the input state. Specifically, each input state has a wavefunction attached to each of the discrete optical mode basis states. These wavefunctions are individually propagated through each section of the network to determine the output.

With a model that describes how the QPNN with QD nonlinearities operates on any input state to produce an output state, we can now train it to perform some task. As discussed in the main text, and as evident from the analysis in the previous section, the scattering in the nonlinear elements distorts the photon wavefunction. This poses a significant threat if the QPNN is trained to perform an operation that must be cascaded, like a CNOT gate for example, since the distortions will stack causing the second CNOT gate to perform a distinct operation from the first CNOT gate. However, if we focus on an operation that is followed by a destructive measurement, like the BSA, then we become much more resistant to these distortions. In fact, they can even play a role in boosting the fidelity as investigated in Fig.~5 of the main text and detailed further in the next section.

\begin{figure}[htb]
\includegraphics[width=\textwidth]{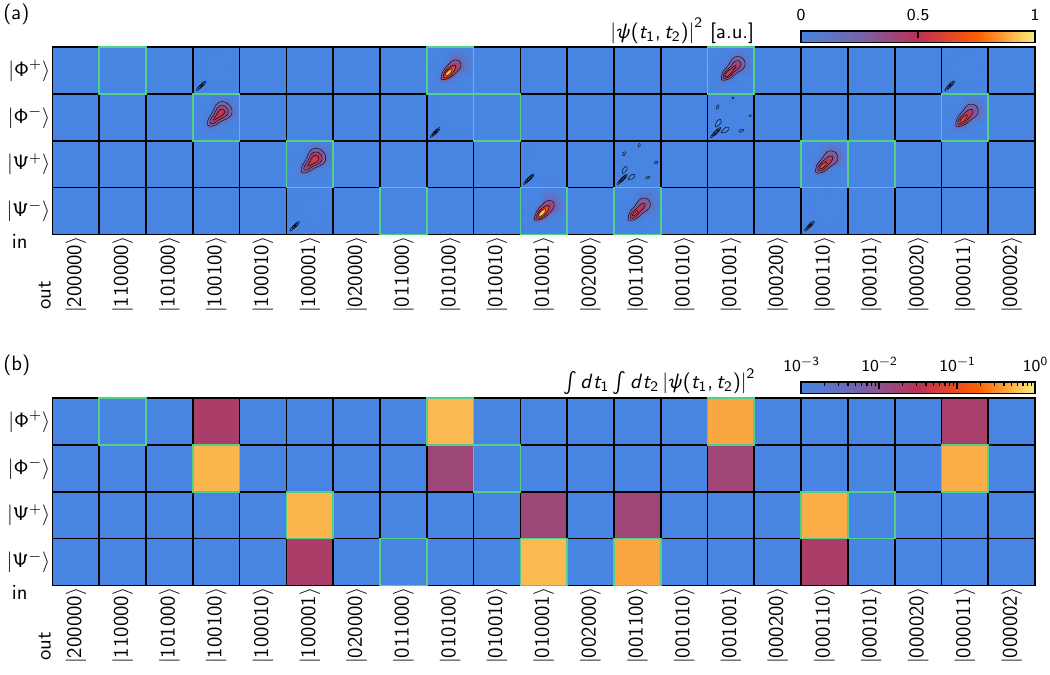}
\centering
\caption{Output states formed by a 4-layer, 6-mode QPNN with QD nonlinearities that is trained to act as a BSA, corresponding to the highest-fidelity result found in all the training trials shown in Fig.~4b of the main text. (a) Two-time probability distributions calculated from the outgoing two-photon wavefunctions attached to each discrete time-bin Fock state (x-axis), for each input Bell-state (y-axis). The wavefunctions are normalized to the maximum value in the entire grid, such that they can all be compared with each other in arbitrary units. For each Bell-state, three measurement outcomes were assigned randomly at the beginning of training, as outlined with green boxes. Contours are drawn in black as a visual aid, only for the wavefunctions that contribute a probability $>0.01$ in the overall state. (b) Probability of measuring each potential outcome (x-axis) for a given input Bell-state (y-axis), with the green boxes outlining successful outcomes as in a.}
\label{fig:bsa-wfs}
\end{figure}

Since we chose to focus on a BSA that is specifically followed by a destructive measurement, we now have a very clear goal in mind: we want to discriminate between all four Bell-states as accurately as possible. In this scenario, it is beneficial to maximize the measurement degrees of freedom that we can reasonably exploit. Specifically, if we assume the use of single-photon detectors that are not necessarily photon-number-resolving (e.g. superconducting nanowire single photon detectors, the standard in the field), then we can distinguish between all measurement outcomes where one photon is measured in mode $m_i$, while the other photon is measured in a distinct mode $m_j$ (i.e. $i \neq j$). Thus, rather than assign one specific measurement outcome to each Bell-state, as is the typical choice for a BSA, we assign an equivalent amount of valid outcomes to each Bell-state randomly at the beginning of each training trial. For instance, consider a $N = 4$ QPNN. Operating on two-photons, there are 10 states in the Fock basis (i.e. 10 combinations of photons allotted to the 4 modes). Out of the 10 states, 4 of these have two photons in the same mode, which cannot be measured accurately using a detector without photon-number-resolution (unable to determine if there are two photons in the same mode or if one of the photons was simply lost). As a result, we are left with 6 measurement outcomes to assign. In this case, we assign just 1 measurement outcome to each Bell state to keep the assignment uniformly distributed. However, if we increase $N$ to 6, there are 15 measurement outcomes to assign, allowing us to assign 3 outcomes to each Bell state (as in the example displayed in Fig.~\ref{fig:bsa-wfs}).
It is likely that the exact assignment does not play a significant role in the QPNN's ability to learn a high-fidelity solution, given the fact that the linear unitaries start with random parameters and that the meshes can learn to perform any arbitrary linear unitary transformation \cite{clements_optimal_2016}. With that said, we randomly assign the outcomes regardless to hopefully ensure that we cover all the permutations over the 100 training runs.

The definitions of fidelity, operational rate, and cost remain the same. Specifically, the fidelity is the chance that the output state is correct given that it is logical. In this case, any measurement outcome that gets assigned becomes logical, because it signals that a specific Bell-state was input. Similarly, the operational rate is computed based on the probability of measuring one of the assigned outcomes. In the data shown in Fig.~5 of the main text, and Fig.~\ref{fig:bsa-wfs} here, all of the probability lies with the assigned measurement outcomes, but the probability that the correct measurement outcomes are actually recorded for a given input Bell state is only 0.96, in the absence of additional time filtering (see Sec.~\ref{sec:filtering}). For this specific case, we see in Fig.~\ref{fig:bsa-wfs} that measurement outcomes $\left\{\left|110000\right\rangle, \left|010100\right\rangle, \left|001001\right\rangle\right\}$ were selected to signify that the Bell-state $\left|\Phi^+\right\rangle$ was input, as an example. The sum of the probability distributions attached to each of these outcomes is what forms $\left|0\right\rangle_\mathrm{out}$ as shown in Fig.~5 of the main text. The same is true for the other outcomes, assigned to the other Bell-states, as well.

With the switch to nonlinear elements based on chiral scattering from QDs, there are new parameters that can be exploited while training the QPNN. Specifically, the pulse width of the photons in time relative to the QD lifetime significantly affects the scattering response, as clear from Eqs.~\ref{eq:tomega} and \ref{eq:somega}. Additionally, the detuning between the photons and the emitter also plays a role. In principle, we could add both of these parameters to every time that a nonlinear element is used in the network, however, it is not currently realistic to consider modulating the lifetime of a QD at high speeds. Thus, while training, we assume that the lifetime of the QD can be adjusted as a parameter, yet there is only one of these parameters for the circuit. Physically, this means that a QD with a specific lifetime would be optimal for the circuit operation, yet it would never need to be actively modulated. In contrast, QD transition frequencies can be readily tuned through the DC Stark shift by applying a bias voltage \cite{Hallett:18, uppu2020scalable_single_photon_QD_4}, the AC Stark shift using optical pumping \cite{le_jeannic_dynamical_2022, Unold:04}, and other means \cite{Grim:19}. Therefore, while we take the QD transition frequency to be constant during each individual layer of nonlinearities, such that it doesn't need to be modulated between consecutive time bins (relatively fast), we allow it to change for each layer of nonlinearities, meaning that it should be modulated between consecutive layers (relatively slow).

%%%%%%%%%%%%%%%%%%%%%%%%%%%%%%%%%%%%%%%%%%%%%%%%%%%%%

\section{Applying Time Filters to the Output of the QPNN BSA} \label{sec:filtering}
In Fig.~5a of the main text, we draw contours at 20\%, 50\%, and 90\% of the maximum of each target probability distribution (i.e. those outlined in green boxes), for each row. These contours serve as an example of how one could design a cross-correlated time filter to isolate specific regions of the probability distribution where there is a high chance of measuring the correct outcome, yet a low chance of measuring an incorrect outcome. As a result, the filter can help boost the overall fidelity of the device at the expense of a slight decrease in operational rate, owing to the fact that some correct photons will be filtered out alongside the incorrect ones.

Experimentally, the simplest approach to implementing this filter is by calculating the cross-correlation between the time tags assigned to each of the two photons measured. This can be done easy with a digital classical computer connected to the time tagger, at the expense of introducing latency between the measurement and the decision on whether the measurement should be thrown out or not. Instead, latencies could be reduced to nanosecond timescales using a dedicated FPGA \cite{photonids}.

%%%%%%%%%%%%%%%%%%%%%%%%%%%%%%%%%%%%%%%%%%%%%%%%%%%%%

% references
% {
\bibliography{ref}
% }